\documentclass[5p]{elsarticle}

\usepackage{lineno,hyperref}
\usepackage{mathtools}
\usepackage{threeparttable}
\usepackage{subfig}
\usepackage[version=3]{mhchem}
\modulolinenumbers[5]

\journal{Journal of Solid State Chemistry}

\biboptions{square,sort&compress}

\begin{document}

\begin{frontmatter}

\title{Synthesis-Dependent Properties of Barlowite and Zn-Substituted Barlowite}

\author[simes,chem]{Rebecca W. ~Smaha\corref{cor1}}
\ead{rsmaha@stanford.edu}
\author[simes,matsci]{Wei He}
\author[simes]{John P. Sheckelton}
\author[simes]{Jiajia Wen}
\author[simes,apphys]{Young S. Lee}
\address[simes]{%
Stanford Institute for Materials and Energy Sciences, SLAC National Accelerator Laboratory, 2575 Sand Hill Road, Menlo Park, California 94025, USA}
\address[chem]{Department of Chemistry, Stanford University, Stanford, California 94305, USA}
\address[matsci]{Department of Materials Science and Engineering, Stanford University, Stanford, California 94305, US}
\address[apphys]{Department of Applied Physics, Stanford University, Stanford, California 94305, USA}

\begin{abstract}
The mineral barlowite, \ce{Cu4(OH)6FBr}, has been the focus of recent attention due to the possibility of substituting the interlayer \ce{Cu}$^{2+}$ site with non-magnetic ions to develop new quantum spin liquid materials. We re-examine previous methods of synthesizing barlowite and describe a novel hydrothermal synthesis method that produces large single crystals of barlowite and Zn-substituted barlowite (\ce{Cu}$_3$\ce{Zn}$_x$\ce{Cu}$_{1-x}$\ce{(OH)6FBr}). The two synthesis techniques yield barlowite with indistinguishable crystal structures and spectroscopic properties at room temperature; however, the magnetic ordering temperatures differ by 4 K and the thermodynamic properties are clearly different. The dependence of properties upon synthetic conditions implies that the defect chemistry of barlowite and related materials is complex and significant. Zn-substituted barlowite exhibits a lack of magnetic order down to \textit{T} = 2 K, characteristic of a quantum spin liquid, and we provide a synthetic route towards producing large crystals suitable for neutron scattering.
\end{abstract}


\end{frontmatter}
\sloppy

\section{Introduction}
Quantum spin liquid (QSL) materials have an exotic magnetic ground state characterized by the spins evading conventional magnetic long-range order down to \textit{T} = 0 K and possessing long-range quantum entanglement.\cite{Balents2010,Norman2016} One way to explain this ground state is as a resonating valence bond state, in which singlets of entangled spins fluctuate over the lattice but never break translational symmetry.\cite{Anderson1973} Through the possibility of obtaining long-range quantum entanglement of spins, a better understanding of the QSL ground state opens avenues to develop materials for topological quantum computing applications.\cite{Ioffe2002}  In addition, investigating QSL candidate materials may have important implications for our understanding of high temperature superconductivity.\cite{Anderson1987,Savary2017}

One of the best experimentally realized QSL candidates is the metal oxyhalide mineral herbertsmithite, \ce{Cu3Zn(OH)6Cl2}.\cite{Braithwaite2004,Shores2005,Han2012,Fu2015} Herbertsmithite has a rhombohedral, layered structure consisting of alternating kagom\'e lattice planes of \ce{Cu}$^{2+}$ ions with layers of nonmagnetic \ce{Zn}$^{2+}$ ions that serve to magnetically isolate the kagom\'e layers. Extreme magnetic frustration can be found when there are competing antiferromagnetic (AFM) interactions between nearest-neighbor S = 1/2 spins on a kagom\'e lattice, which consists of a network of corner-sharing triangles. The physics of herbertsmithite has been studied extensively, but chemical and synthetic limitations have held it back: a small fraction of excess \ce{Cu}$^{2+}$ impurities on the interlayer Zn site results in interlayer magnetic coupling that obscures the intrinsic QSL behavior.\cite{DeVries2012,Han2016b} 

The mineral barlowite\cite{Elliott2014,Han2014}, \ce{Cu4(OH)6FBr}, is another rare example of a material that has an isolated, undistorted S = 1/2 kagom\'e lattice. It contains \ce{Cu}$^{2+}$ ions on its interlayer site, presumably causing it to have a transition to long-range magnetic order at 15 K.\cite{Han2014,Jeschke2015} Barlowite, therefore, is not a QSL material; however, DFT calculations show that substituting the interlayer site with nonmagnetic \ce{Zn}$^{2+}$ or \ce{Mg}$^{2+}$ should suppress the long-range magnetic order and lead to a QSL state.\cite{Guterding2016a,Liu2015a} It has a different coordination environment around the interlayer \ce{Cu}$^{2+}$ (trigonal prismatic as opposed to octahedral in herbertsmithite) and perfect AA stacking of the kagom\'e layers, while herbertsmithite has ABC stacking. It has been predicted that these differences will yield a significantly lower amount of \ce{Cu}$^{2+}$ impurities on the interlayer site in Zn- or Mg-substituted barlowite compared to herbertsmithite, opening up new avenues to study the intrinsic physics of QSL materials.\cite{Liu2015a}

Here, we re-examine the synthesis of barlowite after noting a discrepancy between the morphology of crystals of natural barlowite (described as ``platy" along the \textit{c}-axis\cite{Elliott2014}) and crystals of synthetic barlowite (rods down the \textit{c}-axis).\cite{Han2016,Pasco2018} We present a new method of synthesizing large single crystals of barlowite that are structurally and spectroscopically identical to polycrystalline barlowite at room temperature. However, at low temperatures, the magnetic transition temperature shifts by 4 K. Slight modifications of these two methods produce polycrystalline and single crystalline Zn-substituted barlowite (\ce{Cu}$_3$\ce{Zn}$_x$\ce{Cu}$_{1-x}$\ce{(OH)6FBr}) showing a lack of magnetic order down to \textit{T} = 2 K, consistent with a QSL ground state. This comparison of synthesis methods has implications for past and future studies of related synthetic minerals, especially copper oxysalts produced hydrothermally.  We find that the large dependence of properties on synthetic route suggests that the defect chemistry of copper oxysalts is more complex than previously believed, implying that a true understanding of this class of materials requires careful control over synthesis.

\section{Experimental Details}
\subsection{Materials and Methods}
\ce{Cu2(OH)2CO3} (Alfa, Cu 55\%), \ce{NH4F} (Alfa, 96\%), HBr (Alfa, 48\% wt), \ce{ZnBr2} (BTC, 99.999\%), \ce{CuF2} (BTC, 99.5\%), \ce{LiBr} (Alfa, 99\%), \ce{ZnF2} (Alfa, 99\%), deionized (DI) \ce{H2O} (EMD Millipore), and \ce{D2O} (Aldrich, 99.9\%) were used as purchased. Mid- and near-infrared (IR) measurements were performed on a Thermo Fisher Scientific Nicolet 6700 Fourier transform infrared spectrometer (FTIR) with a Smart Orbit diamond attenuated total reflectance (ATR) accessory. Raman measurements were performed on a Horiba LabRAM Aramis spectrometer with a CCD detector, 1800 grooves/mm grating, and 532 nm laser. DC magnetization measurements were performed on a Quantum Design Physical Properties Measurement System (PPMS) Dynacool from 2 to 350 K under applied fields of 0.005 T, 1.0 T, and 9.0 T. Heat capacity measurements were performed in the PPMS Dynacool on either a pressed single pellet of powder mixed with Ag powder in a 1:2 mass ratio or on a single crystal affixed to a sapphire platform using Apiezon-N grease. 

\subsection{Syntheses}
\textbf{1}: \ce{Cu2(OH)2CO3} (1.5477 g), \ce{NH4F} (0.2593 g), and HBr (0.8 mL) were sealed in a 45 mL PTFE-lined stainless steel autoclave with 36 mL DI \ce{H2O} or \ce{D2O}. This was heated over 3 hours to 175 $^{\circ}$C and held for 72 hours before being cooled to room temperature over 48 hours. The products were recovered by filtration and washed with DI \ce{H2O}, yielding polycrystalline barlowite.

\textbf{1-a}: This was prepared as \textbf{1} above but with the following heating profile: it was heated over 3 hours to 175 $^{\circ}$C and held for 17 days before being cooled to room temperature over 48 hours. 

\textbf{Zn-1}: \ce{Cu2(OH)2CO3} (0.5307 g), \ce{NH4F} (0.0593 g), and \ce{ZnBr2} (0.5405 g) were sealed in a 23 mL PTFE-lined stainless steel autoclave with 10 mL DI \ce{H2O}.  This was heated over 3 hours to 210 $^{\circ}$C and held 24 hours before being cooled to room temperature over 30 hours. The products were recovered by filtration and washed with DI \ce{H2O}, yielding polycrystalline Zn-substituted barlowite.

\textbf{2}: \ce{CuF2} (0.4569 g) and \ce{LiBr} (0.9119 g) were sealed in a 23 mL PTFE-lined stainless steel autoclave with 15 mL DI \ce{H2O}. \textbf{Zn-2}: \ce{CuF2} (0.2742 g), \ce{ZnF2} (0.4653 g), and \ce{LiBr} (1.1724 g) were sealed in a 23 mL PTFE-lined stainless steel autoclave with 15 mL DI \ce{H2O}.  For both, the autoclave was heated over 3 hours to 175 $^{\circ}$C and held for 72 hours, then cooled to 80 $^{\circ}$C over 24 hours. It was held at 80 $^{\circ}$C for 24 hours before being cooled to room temperature over 12 hours.  The products were recovered by filtration and washed with DI \ce{H2O}, yielding barlowite or Zn-substituted barlowite crystals mixed with polycrystalline \ce{LiF}, which was removed by sonication in acetone.

\subsection{X-ray Diffraction}
Single crystal diffraction (SCXRD) experiments were conducted at Beamline 15-ID at the Advanced Photon Source (APS), Argonne National Laboratory, using a Bruker D8 diffractometer equipped with a PILATUS3 X CdTe 1M detector or a Bruker APEXII detector. Datasets were collected at 300 K using a wavelength of 0.41328 {\AA}. The data were integrated and corrected for Lorentz and polarization effects using \textsc{saint} and corrected for absorption effects using \textsc{sadabs}.\cite{BrukerAXSSoftwareInc2016} The structures were solved using intrinsic phasing in \textsc{apex3} and refined using the \textsc{shelxtl}\cite{sheldrick2015} and \textsc{olex2}\cite{Dolomanov2009} software. Hydrogen atoms were inserted at positions of electron density near the oxygen atom and were refined with a fixed bond length and an isotropic thermal parameter 1.5 times that of the attached oxygen atom. Thermal parameters for all other atoms were refined anisotropically. 

High resolution synchrotron powder X-ray diffraction (PXRD) data were collected at 300 K using beamline 11-BM at the APS using a wavelength of 0.412728 {\AA}. Samples were measured in kapton capillaries; crystalline samples were crushed into a powder.  Rietveld refinements were performed using \textsc{gsas-II}.\cite{toby2013}  Atomic coordinates and isotropic atomic displacement parameters were refined for each atom. Following results from SCXRD, the site occupancy of the interlayer Cu or Zn was fixed at 0.3333; however, for \textbf{Zn-1} the Zn refined to nearly octahedral, so it was placed on the octahedral site with an occupancy of 1. Hydrogen was excluded.

\section{Results and discussion}

\subsection{Synthesis}
Attempts to replicate the reported synthesis of barlowite\cite{Han2014} were stymied by its use of \ce{HBrO4}, an unstable\cite{Appelman1969} and commercially unavailable reagent. Replacing \ce{HBrO4} with \ce{HBr} yielded crystals an order of magnitude smaller than those reported previously\cite{Jeschke2015,Han2016} and too small for neutron scattering experiments. Thus, we developed two alternate synthetic routes to produce barlowite and Zn-substituted barlowite:

\textbf{Method 1:}

\textbf{1a:} \ce{2Cu2(OH)2CO3 + NH4F + HBr + H2O ->} 

\ce{Cu4(OH)6FBr + 2CO2 + NH3}

\textbf{1b:} \ce{3Cu2(OH)2CO3 + 2ZnBr2 + 2NH4F + H2O ->}

\ce{2Cu3Zn(OH)6FBr + 3CO2 + 2NH3}

\textbf{Method 2:}

\textbf{2a:} \ce{4CuF2 + 7LiBr + 6H2O -> Cu4(OH)6FBr +}

\ce{7LiF + 6HBr}

\textbf{2b:} \ce{3CuF2 + ZnF2 + 7LiBr + 6H2O -> }

\ce{Cu3Zn(OH)6FBr + 7LiF + 6HBr}

Method 1 produces polycrystalline barlowite mixed with small crystals (up to 0.5 mm, \textbf{1}) or polycrystalline Zn-substituted barlowite (no crystals, \textbf{Zn-1}). Method 1 is similar to the first reported syntheses of barlowite\cite{Han2014} and Zn-substituted barlowite.\cite{Feng2017a} However, we utilize slightly different reagents (\ce{HBr} instead of \ce{HBrO4} in Method 1a and no \ce{CuBr2} in Method 1b, and we have optimized the stoichiometry and temperature profile. The higher temperature used here for barlowite (175 $^{\circ}$C, Method 1a) yields 0.5 mm crystals more efficiently than synthesis at 120 $^{\circ}$C.\cite{Pasco2018} However, neither this method nor any literature report on Zn-substituted barlowite\cite{Feng2017,Feng2017a} produces crystals, which will hinder future neutron studies of its possible QSL ground state.

Method 2 produces large crystals (up to 2 mm) of both barlowite (\textbf{2}) and Zn-substituted barlowite (\textbf{Zn-2}). Method 2a is an entirely novel route for the growth of single crystals of barlowite, and they must be mechanically separated from the \ce{LiF} byproduct.  The preferential formation and stability of \ce{LiF} will aid the growth of high-quality barlowite crystals. It is modified to produce Zn-substituted barlowite (Method 2b) with the addition of a large excess of \ce{ZnF2} and correspondingly increasing the \ce{LiBr} stoichiometry. 

As measured by inductively coupled plasma atomic emission spectroscopy (ICP-AES), the Zn content of polycrystalline \textbf{Zn-1} is 0.95, and the Zn content of \textbf{Zn-2} averaged over several crystals is 0.33.  These were produced using 1.5 equivalents of \ce{ZnBr2} and 5 equivalents of \ce{ZnF2}, respectively, and the difference can be attributed to the nearly five orders of magnitude higher solubility of \ce{ZnBr2} in water compared to \ce{ZnF2}.

Both methods utilize the moderate temperature and pressure range accessible to PTFE-lined autoclaves.  PTFE is essential given the presence of fluorine in the reaction; attempting to synthesize single crystalline barlowite in a quartz tube such as for herbertsmithite\cite{Chu2011,Han2011} is futile since \ce{F-} is expected to etch the quartz before forming barlowite. The two synthesis methods produce barlowite crystals with different morphologies. Method 1 crystals (Figure \ref{fgr:Struc}C) grow as small hexagonal rods whose long axis is the \textit{c}-axis, similar to those reported in the literature,\cite{Han2016,Pasco2018} while Method 2 crystals (Figure \ref{fgr:Struc}D) grow as larger hexagonal plates flattened along the \textit{c}-axis.  Naturally-occurring barlowite crystals were described as `platy' and thus are likely more similar to those produced by Method 2.\cite{Elliott2014} 

\subsection{Structure and Composition}
The reported room temperature crystal structure as solved via single crystal X-ray diffraction (SCXRD) structure in space group \textit{P}6\textsubscript{3}/\textit{mmc} (No. 194)\cite{Han2014} agrees well with that of naturally-occurring barlowite.\cite{Elliott2014}  Recent reports disagree on whether the crystal structure is hexagonal\cite{Feng2017a} or orthorhombic\cite{Pasco2018} at room temperature. Single crystal X-ray diffraction measurements at beamline 15-ID at the APS at \textit{T} = 300 K performed on crystals of \textbf{1}, \textbf{2}, and \textbf{Zn-2} showed no signs of symmetry lowering or pseudomerohedral twinning, so we assign the structure as hexagonal space group \textit{P}6\textsubscript{3}/\textit{mmc}. Lattice parameters and refinement details for the SCXRD structures can be found in Table S1,\cite{supp} and extracted bond distances can be found in Table S2.

\begin{figure*}[hbt!]
\includegraphics[width=17cm]{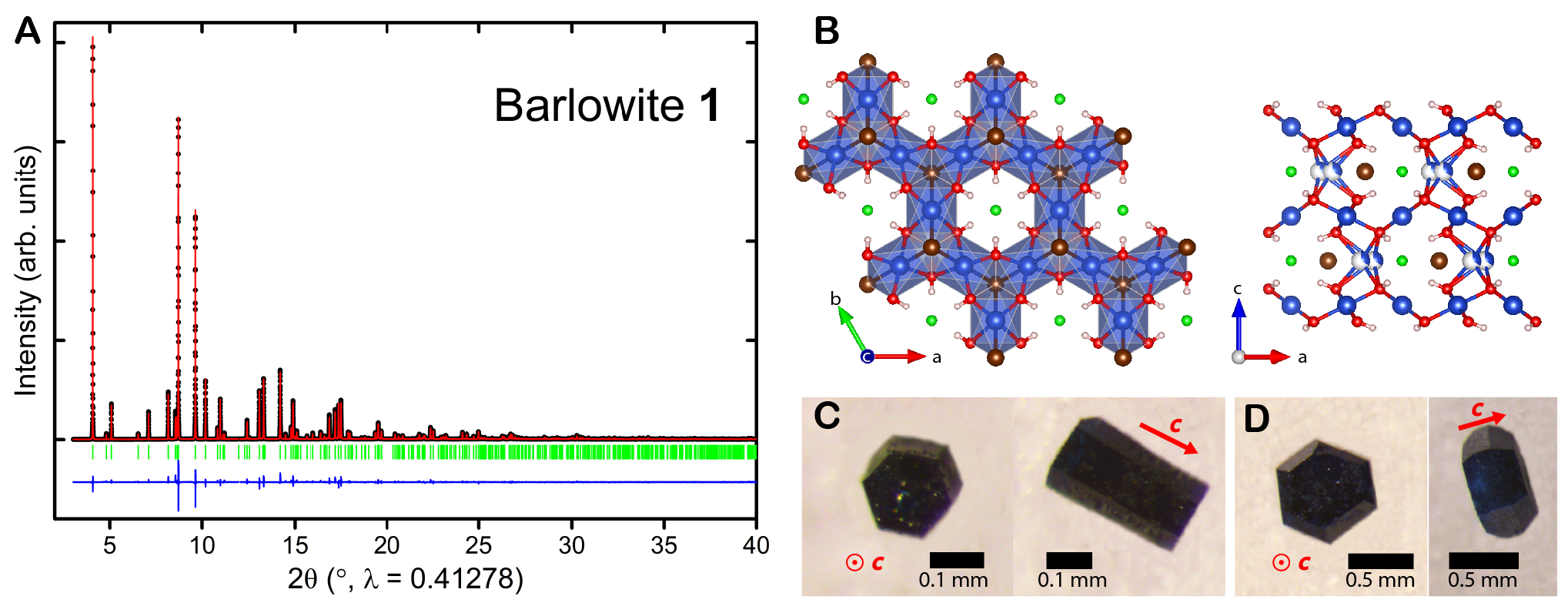}
\caption{A) Representative Rietveld refinement of synchrotron PXRD data of \textbf{1} at  \textit{T} = 300 K in space group \textit{P}6\textsubscript{3}/\textit{mmc}.  Observed (black), calculated (red), and difference (blue) plots are shown, and Bragg reflections are indicated by green tick marks. B) Structure of barlowite as determined by single crystal diffraction, visualized in VESTA.\cite{Momma2011} Blue, brown, red, green, and white spheres represent Cu, Br, O, F, and H atoms, respectively. C) \textbf{1} crystal and D) \textbf{2} crystal.}
\label{fgr:Struc}
\end{figure*}

The structure of barlowite is depicted in Figure \ref{fgr:Struc}B, showing the kagom\'e plane of highly Jahn-Teller (4+2)-distorted \ce{CuO4Br2} octahedra and the coordination environment around the interlayer Cu site. This interlayer Cu is disordered over three symmetry-equivalent sites and has trigonal prismatic coordination, isostructural to claringbullite.\cite{Burns1995a} While rare, trigonal prismatic \ce{Cu}$^{2+}$ occurs in several other copper oxysalt minerals besides claringbullite, including buttgenbachite\cite{Fanfani1973} and connellite.\cite{Mclean1972} Each symmetry-equivalent interlayer site has four short Cu-O bonds ($\sim$2.0 {\AA}) to the nearer oxygens and two long Cu-O bonds ($\sim$2.4 {\AA}). 

The near-identical values of ionic radii, X-ray scattering factors, and neutron scattering factors of Cu and Zn make accurately distinguishing site occupancies of these two elements using diffraction techniques extremely difficult. \ce{Zn}$^{2+}$ is not Jahn-Teller active, and therefore substituting onto the kagom\'e site is energetically unfavorable. Having up to 15\% excess \ce{Cu}$^{2+}$ ions on the Zn interlayer site is possible and has been shown in herbertsmithite using anomalous scattering measurements.\cite{Freedman2010}  In the absence of anomalous diffraction measurements to definitively determine the Cu:Zn ratio and site occupancies in Zn-substituted barlowite, we fix the Zn to substitute on the interlayer site in both single crystal and Rietveld refinements. Following ICP-AES results, we fix the Cu:Zn ratio to be 3:1 and 3.67:0.33 in \textbf{Zn-1} and \textbf{Zn-2}, respectively.

High resolution synchrotron PXRD datasets were collected at beamline 11-BM at the APS at \textit{T} = 300 K for barlowite and Zn-substituted barlowite synthesized using both methods.  A representative Rietveld refinement in space group \textit{P}6\textsubscript{3}/\textit{mmc} for \textbf{1} is shown in Fig. \ref{fgr:Struc}A, and crystallographic data are tabulated in Table S3.  The remaining Rietveld refinements are shown in Figures S1--S3. The refinements show that the two methods produce crystallographically identical samples and support the assignment of hexagonal symmetry. 

Selected bond distances are shown in Table \ref{tbl:bond300K}; the structural effect of Zn substitution is visible as a shift in the triplicated, disordered interlayer site (Cu2 or Zn1). The length of one side of the triangle formed by this disorder (Cu2--Cu2) in both \textbf{1} and \textbf{2} is approximately 0.74 {\AA}, while in \textbf{Zn-1} the Zn moves to the center of this site and becomes octahedral.  Since \ce{Zn}$^{2+}$ is not Jahn-Teller active, it is closer to the center of an octahedron instead of at the extremes of a trigonal prismatic geometry, which is more likely for the Jahn-Teller active \ce{Cu}$^{2+}$.\cite{Burns1996}  The intermediate bond length of \textbf{Zn-2} (0.59 {\AA}) reflects the lower amount of \ce{Zn}$^{2+}$ that has been substituted onto the interlayer site compared to \textbf{Zn-1}.  These trends are corroborated by bond distances extracted from single crystal refinements (Table S2).
 
 \begin{table*}[htb!]
    \centering
        \caption{Selected bond distances extracted from Rietveld refinements of barlowite and Zn-substituted barlowite,  \textit{T} = 300 K. For \textbf{Zn-2}, the interlayer metal position (here called Zn1) is occupied by 0.67 Cu and 0.33 Zn.}
    \begin{tabular}{|c|c|c||c|c|c|}
    \hline
    \textbf{atoms} & Barlowite \textbf{1} & Barlowite \textbf{2} & \textbf{atoms} & \textbf{Zn-1} & \textbf{Zn-2}\\
    \hline 
    Cu1--Cu1 & 3.33871(0) {\AA} &  3.34001(0) {\AA} & Cu1--Cu1 & 3.33796(0)  {\AA} &  3.33812(0) {\AA}\\
    Cu1--O1 & 1.9635(11) {\AA} &  1.9755(9) {\AA}& Cu1--O1 & 1.9758(6) {\AA}& 1.9594(11) {\AA}\\
    Cu1--Br1 & 3.01999(0) {\AA}& 3.02027(0) {\AA}&  Cu1--Br1 & 3.02416(0) {\AA}& 3.02153(0) {\AA}\\
    Cu2--O1 (1) & 2.0030(18) {\AA}&  1.9861(17) {\AA}& Zn1--O1 (1) & 2.1076(9) {\AA}&  2.0294(13) {\AA}\\
    Cu2--O1 (2) & 2.0030(17) {\AA}&  1.9861(16) {\AA}& Zn1--O1 (2) & n/a &  2.0293(12) {\AA}\\
    Cu2--O1 (3) & 2.4455(12) {\AA}& 2.4266(10) {\AA}& Zn1--O1 (3) & n/a & 2.3860(12) {\AA}\\
    Cu2--Cu2 & 0.74153(0) {\AA}& 0.73180(0) {\AA}& Zn1--Zn1 & 0 {\AA}& 0.59253(0) {\AA}\\  
    \hline
    \end{tabular}
    \label{tbl:bond300K}
\end{table*}
 
 \subsection{Fourier Transform Infrared and Raman Spectroscopy}
\begin{figure}[htb!]
\includegraphics{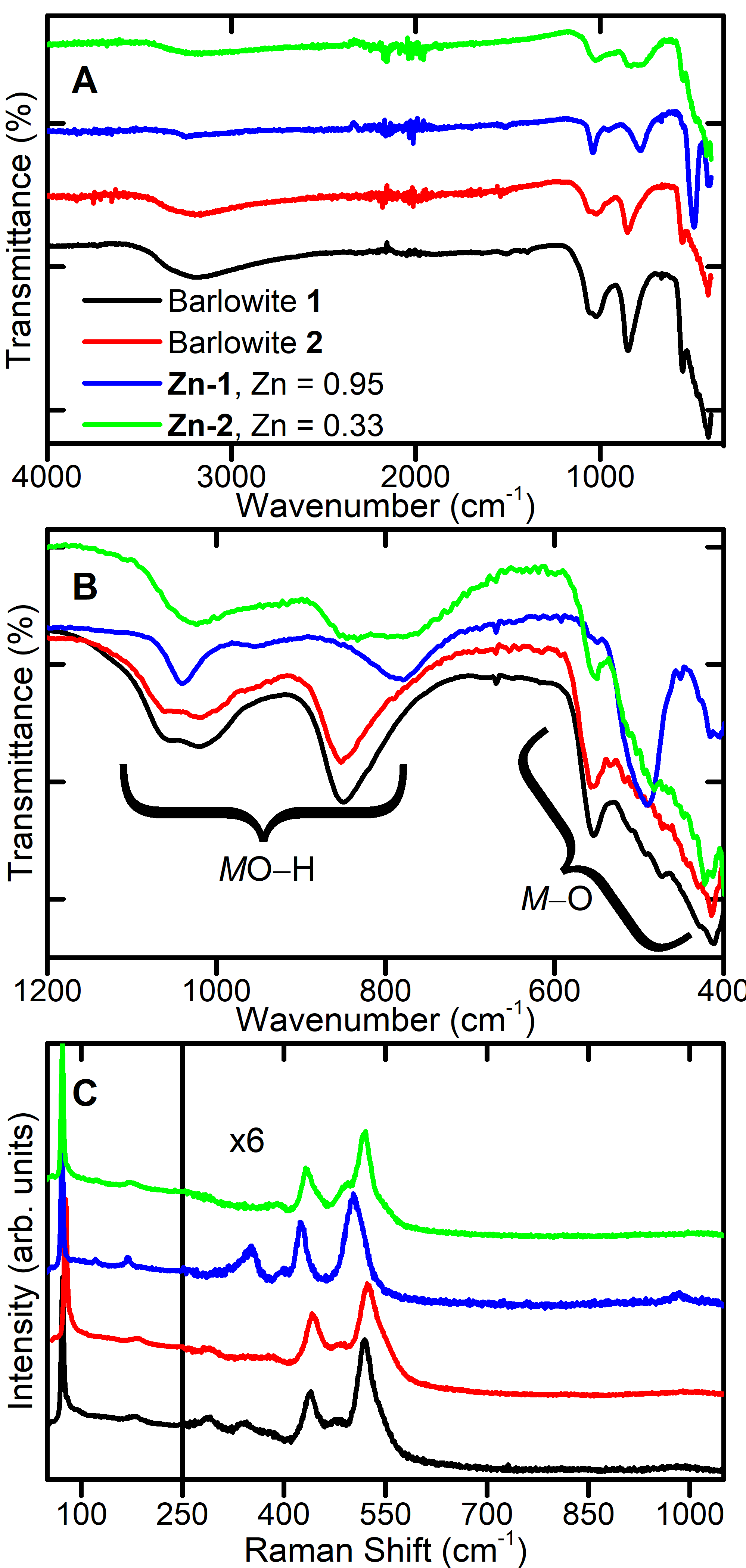}
\caption{A), B) FTIR and C) Raman spectra ($\lambda$ = 532 nm) of barlowite and Zn-substituted barlowite comparing the two synthesis methods.}
    \label{fgr:IR}
\end{figure}

 Polycrystalline (Method 1) and crushed single crystals (Method 2) of barlowite and Zn-substituted barlowite were examined using attenuated total reflectance (ATR) Fourier Transform Infrared Spectroscopy (FTIR) at room temperature, shown in Figure \ref{fgr:IR}A. The spectra have a broad band of O--H stretches centered at $\sim$3100 cm$^{-1}$. The modes in the 700--1060 cm$^{-1}$ range are assigned to CuO--H and ZnO--H deformations, while the modes between 400--700 cm$^{-1}$ are likely due to Cu--O or Zn--O stretches (Figure \ref{fgr:IR}B).\cite{Braithwaite2004,Schuiskii2013,Sithole2012}  The modes shift slightly between barlowite and Zn-substituted barlowite, reflecting the mixture of Cu and Zn in the \textit{M}O--H and \textit{M}--O regions. Both barlowite samples have modes at 1056 and 1020 cm$^{-1}$ and a stronger mode at 850 cm$^{-1}$, which shift to 1040 and 782 cm$^{-1}$ when nearly a full equivalent of Zn is substituted into the structure (\textbf{Zn-1}). As \textbf{Zn-2} has much less Zn (0.33 compared to 0.95), its spectra shows a combination of the two end points, resulting in broad modes at 1020, 845, and 780 cm$^{-1}$.
 
In the \textit{M}--O region (400--700 cm$^{-1}$), the mode at 553  cm$^{-1}$ is found in all samples. The relative strength of the mode at 490 cm$^{-1}$ in \textbf{Zn-1}, which is seen only as a weak shoulder in \textbf{Zn-2}, may be due to the much higher amount of Zn present in \textbf{Zn-1}. These determinations are supported by a shift in the O--H band and in the modes in the 700--1060 cm$^{-1}$ region upon deuteration, whereas features below 600 cm$^{-1}$ are unaffected (see Figure S4). While the presence of fluorine may affect the stretching frequencies compared to herbertsmithite, the recent assignment of all modes below 1100 cm$^{-1}$ as F--H or F--D stretches\cite{Pasco2018} is not supported by other work on the spectroscopy of Cu- or Zn-containing hydroxy minerals\cite{Braithwaite2004,Schuiskii2013,Sithole2012} and does not explain the shift in modes within the 400-1060 cm$^{-1}$ region between barlowite and Zn-substituted barlowite. While slight spectral differences between barlowite and Zn-substituted barlowite are expected due to the substitution of Zn, both synthetic routes produce spectroscopically equivalent samples.

 Raman spectroscopy at room temperature was performed on polycrystalline (Method 1) and single crystalline (Method 2) barlowite and Zn-substituted barlowite with a laser excitation of 532 nm (Figure \ref{fgr:IR}C). There is a strong mode at approximately 75 cm$^{-1}$ and weaker modes at 185 and 430 cm$^{-1}$ in all samples, with minor shifts. Both barlowite samples have a relatively strong mode at 520 cm$^{-1}$, while in \textbf{Zn-1} it shifts to 500 cm$^{-1}$.  \textbf{Zn-2} contains both modes, reflecting its mixture of Zn and Cu on the interlayer site. While the spectra show good agreement between the two barlowite samples, some differences exist between the Zn-substituted barlowite samples. There are additional peaks at 350 and 985 cm$^{-1}$ in \textbf{Zn-1}; we hypothesize that these may due to the larger amount of Zn present compared to \textbf{Zn-2}. 

 \subsection{Magnetic Susceptibility}
\begin{figure}[htb!]
\includegraphics{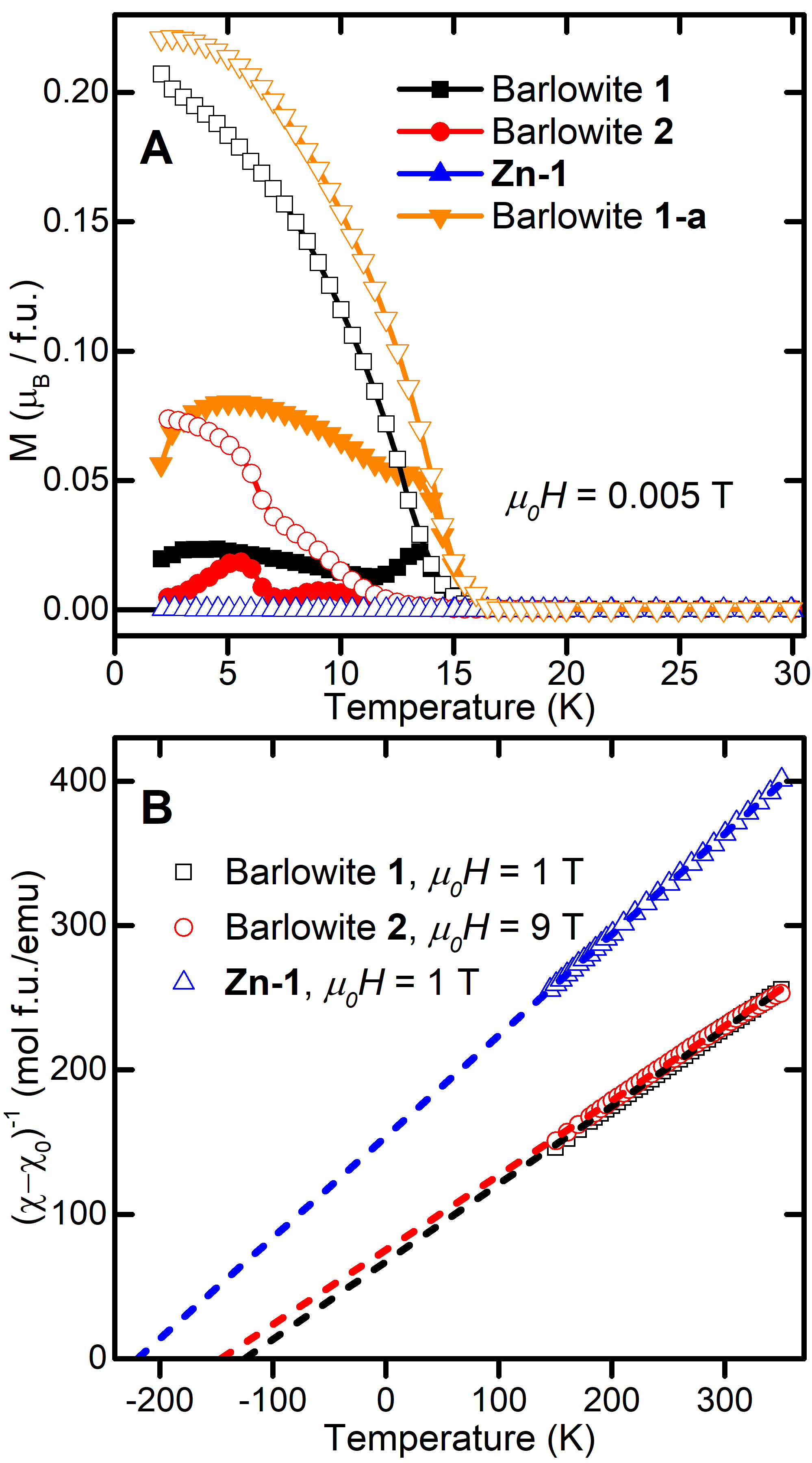}
  \caption{A) ZFC (closed symbols) and FC (open symbols) magnetization of barlowite and \textbf{Zn-1} measured in an applied field of $\mu_0$\textit{H} = 0.005 T. B) Inverse susceptibility data and Curie-Weiss fits extrapolated to the Weiss temperature.}
    \label{fgr:susc}
\end{figure}

Low-field ($\mu_0$\textit{H} = 0.005 T) zero field cooled (ZFC) and field cooled (FC) DC susceptibility measurements were performed on polycrystalline \textbf{1} and a collection of single crystals of \textbf{2} (Figure \ref{fgr:susc}A).  \textbf{1} has a steep onset at \textit{T$_N$} = 15 K, consistent with previous reports.\cite{Han2014,Jeschke2015} However, \textbf{2} has a gradual onset at \textit{T$_N$} = 11 K as well as a second transition at \textit{T} = 6 K.  The higher temperature transition in both samples appears to have some ferromagnetic (FM) character, as indicated by the bifurcation between the FC and ZFC measurements.\cite{Domenicali1950} The magnitude of the magnetization of \textbf{1} is approximately twice that of \textbf{2} at \textit{T} = 2 K. Barlowite synthesized by Method 1 using a longer dwelling time at 175 $^{\circ}$C (17 days instead of 3 days; denoted \textbf{1-a}) exhibits a higher ordering temperature (\textit{T$_N$} $\approx$ 16.5 K) as well as a different response between \textit{T} = 2 and 15 K yielding a larger magnitude of the magnetization. The difference in low temperature magnetic properties between materials with seemingly identical room temperature structures and spectroscopic properties calls into question the validity of comparing samples reported in the literature using different synthesis methods.  

Curie-Weiss fits of high temperature (\textit{T} = 180--350 K) inverse susceptibility data of barlowite (Figure \ref{fgr:susc}B) reveal slight differences between the two methods. A diamagnetic correction  $\chi_0$ = -0.00025 emu/mol was obtained from an initial fit for \textbf{1}, and this value was fixed for all subsequent Curie-Weiss fits, with the assumption that the difference in diamagnetic correction between the samples is negligible. As shown in Table \ref{tbl:CW}, there is good agreement between the values of the effective magnetic moment ($\mu_{eff}$) for each \ce{Cu}$^{2+}$ ion and the \textit{g} factor (assuming S = 1/2) for all barlowite samples. The values of the molar Curie constant (C) are slightly different but both reasonable for \ce{Cu}$^{2+}$. The extracted Weiss temperatures ($\Theta$) are quite large---indicating strong antiferromagnetic (AFM) interactions---and both show good agreement with the reported value ($\Theta$ = -136(10) K).\cite{Han2014} The deviations in susceptibility from the Curie-Weiss fit below 180 K and the large ratios between the Weiss temperature and the N\'eel transition temperature indicate a high degree of magnetic frustration, yielding a frustration index \textit{f} greater than 8 for all barlowite samples.\cite{Ramirez1994} 

 The Curie-Weiss fit of \textbf{1-a} (shown in Figure S5) yields a larger Weiss temperature than \textbf{1}, suggesting that a longer synthesis dwelling time affects the magnetism through a process akin to annealing, allowing defects within the structure to move to a more energetically favorable position.  Barlowite synthesized starting with \ce{CuF2} (\textbf{2}) affords a third set of magnetic properties---it yields the highest Weiss temperature and highest frustration index \textit{f} but lowest FM transition temperature. The differences between \textbf{1}, \textbf{1-a}, and \textbf{2}, which give identical PXRD patterns at room temperature, imply that the magnetism is disproportionately affected by subtle differences in defects controlled by synthesis conditions.
 
 \begin{table*}[htb!]
  \caption{Curie-Weiss parameters fit from \textit{T} = 180--350 K. $\chi_0$ = -0.00025 emu/mol for samples synthesized in this work.}
  \label{tbl:CW}
  \begin{tabular}{|p{3.5cm}|p{2.3cm}|p{1.8cm}|p{1.8cm}|p{1.8cm}|p{1.8cm}|}
  \hline
      & Barlowite\cite{Han2014} & \textbf{1} & \textbf{1-a} & \textbf{2} & \textbf{Zn-1} \\
    \hline 
C (K$\cdot$emu/mol) & -- & 1.860(3) & 1.863(9) & 1.937(11) & 1.431(4) \\
$\Theta$ (K) & -136(10) & -125(1) & -134(2) &-146(2) & -220(1) \\
$\mu_{eff}$ ($\mu_{B}$) & -- & 1.928(2) & 1.929(5) & 1.967(5) & 1.953(3)\\
 \textit{g} factor & 2.27 & 2.226(2) & 2.228(5) & 2.272(6) &  2.255(3)\\
 Frustration index \textit{f} & 9.1(9) & 8.3(6) & 8.1(5) & 13.3(12) & infinite\\
 \hline
  \end{tabular} 
\end{table*}

Polycrystalline samples of Zn-substituted barlowite without a magnetic transition down to \textit{T} = 2 K have been reported,\cite{Feng2017,Feng2017a} although ours is the first report of single crystals. Low-field ZFC and FC DC susceptibility measurements on polycrystalline \textbf{Zn-1} are also shown in Figure \ref{fgr:susc}A. It shows no signs of magnetic order down to \textit{T} = 2 K, suggesting a QSL ground state.  Our syntheses of \textbf{Zn-2} have produced materials with 33\% substitution of \ce{Zn}$^{2+}$ on the interlayer site, which suppresses the ordering temperature to \textit{T$_N$} = 4 K. Curie-Weiss fits to the high temperature (180-350 K) inverse susceptibility data of \textbf{Zn-1} using a diamagnetic correction  $\chi_0$ = -0.00025 emu/mol are shown in Figure \ref{fgr:susc}B, and the extracted values are summarized in Table \ref{tbl:CW}. The molar Curie constant  is 76.9\% of that of \textbf{1}, in good agreement with the theoretical value of 75\% for a fully-substituted Zn-barlowite with three magnetic \ce{Cu}$^{2+}$ ions to barlowite's four. The Weiss temperature $\Theta$ = -220(1) K is more negative than that of barlowite. The less negative values found for barlowite are likely due to a ferromagnetic component from magnetic interactions related to the interlayer Cu. 

\subsection{Heat Capacity}
\begin{figure}[htb!]
\includegraphics{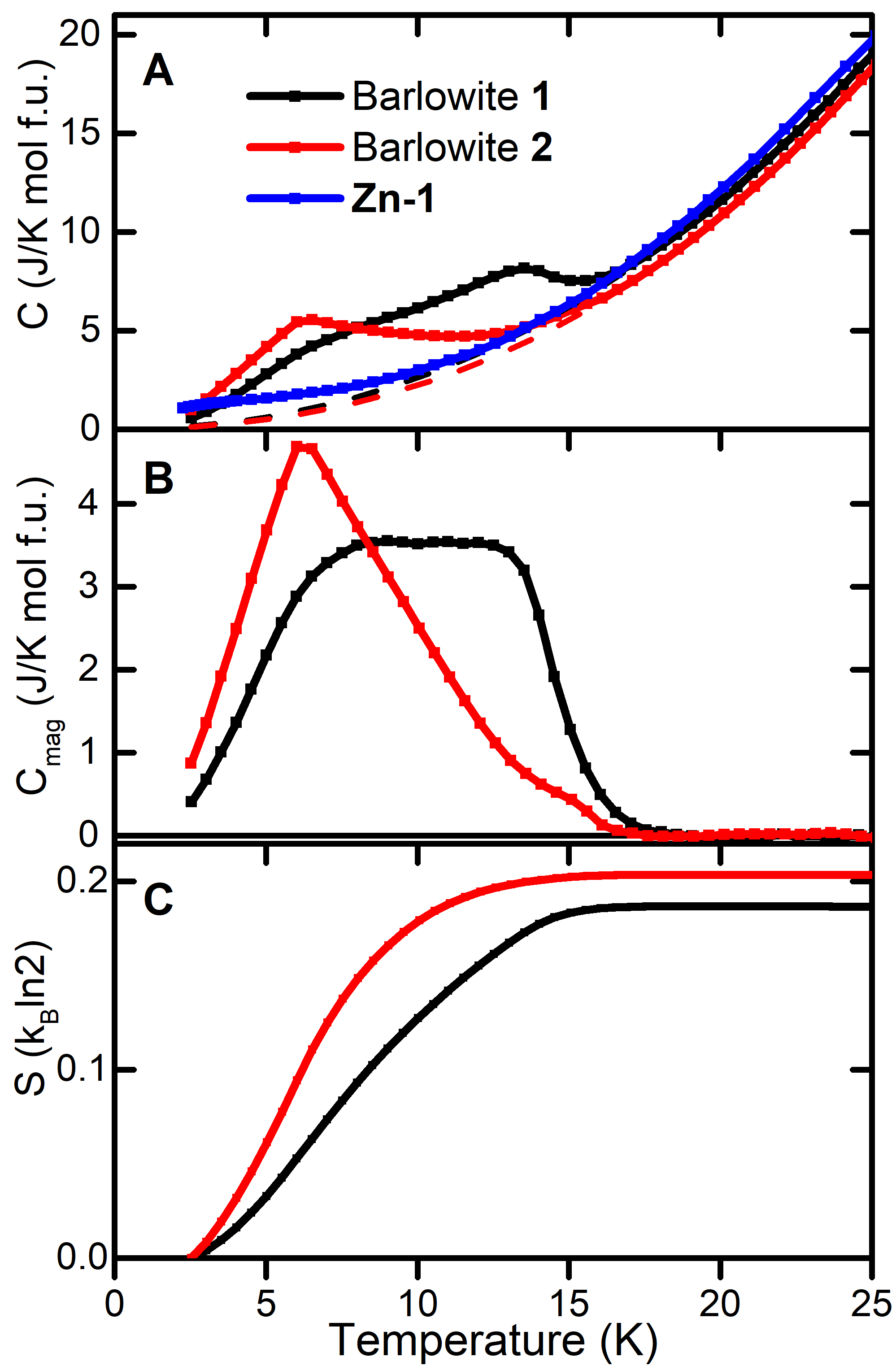}
\caption{Heat capacity (HC) measurements on \textbf{1} and \textbf{Zn-1} (pressed powder mixed with Ag) and \textbf{2} (single crystal). A) Molar HC; the dashed lines indicate \textit{C$_{bg}$} for each sample.  B) \textit{C$_{mag}$} calculated by subtracting the background from the molar HC. C) Magnetic entropy normalized as a fraction of the total value per Cu.}
\label{fgr:HC}
\end{figure}

Heat capacity (HC) measurements were performed from \textit{T} = 2.5--25 K on pressed pellets of polycrystalline \textbf{1} and \textbf{Zn-1} and a 2.0 mg single crystal of \textbf{2}. The powders were mixed with Ag powder to improve thermal conductivity; the contribution of Ag was removed by measuring and subtracting pure Ag. The two barlowite samples show markedly different behavior below 20 K, corroborating the magnetization data.  In the molar HC data (\textit{C}, Figure \ref{fgr:HC}A), \textbf{1} exhibits a broad, asymmetric feature peaking at 13.5 K while \textbf{2} has a narrower peak centered at 6.5 K.  \textbf{Zn-1} does not exhibit a magnetic transition or any other magnetic feature down to \textit{T} = 2.5 K, consistent with a QSL ground state, and its HC and that of \textbf{Zn-2} is a topic of ongoing research and will be discussed further in future work. The small displacement between the curves above 20 K can be ascribed to the uncertainty in the mass normalization. For both barlowite samples, the background was fit to a third-degree polynomial \textit{C$_{bg}$ = aT$^2$ + bT$^3$} between \textit{T} = 18--25 K, following a previous analysis.\cite{Han2014}  One expects the cubic term (\textit{bT$^3$}) to derive from crystal lattice contribution; however, an additional quadratic term (\textit{aT$^2$}) improved the fit in this range significantly (which is likely related to an intrinsic contribution from the kagom\'e spins). Since we aim to examine the anomalies in the HC related to the magnetic transitions of \textbf{1} and \textbf{2} and directly compare them to that reported by Han et al.,\cite{Han2014} we treat this empirical polynomial fit as a background in this discussion.

\textit{C$_{bg}$} was subtracted from \textit{C} to obtain the HC related to the magnetic transition (\textit{C$_{mag}$}, Figure \ref{fgr:HC}B); \textit{C$_{mag}$}/T was integrated from 2.5--25 K to determine the entropy released by this transition (S, Figure \ref{fgr:HC}C). The \textit{C$_{mag}$} of \textbf{1} has a sharp onset and a plateau between 7--14 K. While the onset temperature for \textbf{1} ($\sim$15 K) is the same as that reported by Han et al.\cite{Han2014} for barlowite synthesized using similar precursors to our Method 1a, \textbf{1} has a much broader and flatter plateau down to $\sim$7 K than the reported barlowite. This is due to the different relative intensity of a shoulder at $\sim$7 K: it is weaker than the 15 K feature in the Han et al. sample but equally as strong as the 15 K feature in \textbf{1}.  That subtle differences occur in the HC between these two samples--whose synthesis methods are much closer than comparing \textbf{1} and \textbf{2}--further reveals the dependence of the physical properties upon synthesis condition.

Compared to \textbf{1} and the Han et al. sample, \textbf{2} has a broader onset and a sharp peak at 6 K, potentially correlated to the two transitions seen in the magnetization data. The magnetic entropy per Cu released by the transition to long-range magnetic order is plotted in Figure \ref{fgr:HC}C and is qualitatively similar to the literature report.\cite{Han2014} Barlowite \textbf{2} has a steeper onset at lower temperature and reaches a higher value than \textbf{1}. For both samples, the magnetic entropy is significantly lower than the expected value if all of the Cu spins become ordered at the transition. This may be intrinsic and due to the formation of dynamic spin correlations, as suggested in previous work.\cite{Han2014}  The plateaus in the entropy above 20 K are artifacts of the background subtraction. 

\subsection{Discussion}
Overall, the results described above point to a complex picture of materials issues affecting the structure and properties of barlowite.  The two versions of barlowite synthesized here differ significantly in their magnetic and thermodynamic properties. The two synthesis methods utilize different sources of \ce{Cu}$^{2+}$ ions, and we posit that this leads to distinct reaction mechanisms.  When barlowite is synthesized from \ce{CuF2} (Method 2), the Cu-F bonds must break so that the Cu-O and Cu-Br bonds can form, while each \ce{Cu}$^{2+}$ ion in \ce{Cu2(OH)2CO3} (Method 1) already has four Cu-O bonds. These distinct reaction pathways and transition states could make different types of lattice defects more energetically favorable in each variant of barlowite. Defects in natural minerals are common, depending greatly on the environment during crystal formation, and can affect magnetic and physical properties dramatically.\cite{Schock1985,Hobbs1984} The relatively mild temperature and pressure conditions under which barlowite and other Cu-containing oxysalt minerals crystallize may permit defects such as oxygen or copper vacancies to form, and small differences in these environments seem to have a large effect upon the resulting material. The family of copper oxysalt minerals contains a wide diversity of stable coordination environments for its \ce{Cu}$^{2+}$ ions;\cite{Burns1995} small divergences are thus not likely to destabilize the overall structure. Given that the two variants of barlowite are indistinguishable crystallographically at room temperature, either the difference in defects alone is enough to affect the physical properties, which is plausible given the effect of Cu/Zn site mixing upon the magnetic properties of herbertsmithite,\cite{DeVries2012,Han2016b} or there may be some difference in low temperature structure engendered by the different reaction pathway.  

Materials synthesis provides the ability to optimize growth conditions to make a sample as pure as possible in order to measure the intrinsic properties. In the case of barlowite, we present two options and must now determine which is the ``true" or ``best" barlowite. In some areas, such as semiconductor processing, the ``best" materials are those that are the most defect-free.  As direct measurements of the defect levels in barlowite via transport measurements are complicated by its insulating nature, other metrics must be considered. Potential evaluation criteria for barlowite, taking into consideration that it is the parent compound to a quantum spin liquid material characterized by the lack of long-range magnetic order, could include the temperature of magnetic ordering transitions or the ease of synthesizing crystals suitable for neutron scattering experiments. However, more work must be done to investigate the low-temperature properties and rich physics of this system; perhaps both variants of barlowite will shed light upon the fundamental excitations of the frustrated antiferromagnetic \ce{Cu}$^{2+}$ kagom\'e lattice.

\section{Conclusion}
We re-examine the reported synthesis of barlowite (\ce{Cu4(OH)6FBr}) and Zn-substituted barlowite (\ce{Cu}$_3$\ce{Zn}$_x$\ce{Cu}$_{1-x}$\ce{(OH)6FBr}), and we present a novel method that yields large single crystals. These two synthetic routes yield barlowite and Zn-substituted barlowite with the same structure and FTIR and Raman spectra at room temperature.  However, the magnetic properties of barlowite produced via these two methods diverge at low temperatures: Method 1 barlowite has a transition to long-range magnetic order at \textit{T$_N$} = 15 K, matching previously reported magnetic properties, while Method 2 barlowite has a transition at \textit{T$_N$} = 11 K and a second transition at \textit{T} = 6 K.  The heat capacity at low temperature also differs significantly between Method 1 and Method 2 barlowite. Given that both methods produce structurally equivalent materials, this difference raises questions about the role that synthesis-related defects play in the physical properties of barlowite and similar materials.  

Modifying the two synthesis methods yields Zn-substituted barlowite: Rietveld refinements, ICP-AES analysis, and magnetic data support the successful introduction of Zn into the structure.  Method 1 produces polycrystalline Zn-substituted barlowite with a formula determined by ICP-AES to be \ce{Cu$_{3.05}$Zn$_{0.95}$(OH)6FBr}; it does not order magnetically down to \textit{T} = 2 K and shows highly frustrated behavior consistent with that of a quantum spin liquid material. While Zn-substituted barlowite synthesized via Method 2 orders at \textit{T$_N$} =  4 K, consistent with its lower Zn content of \ce{Cu$_{3.67}$Zn$_{0.33}$(OH)6FBr}, it produces the first single crystals of Zn-substituted barlowite. This provides a synthetic route towards the production of large single crystals suitable for neutron scattering.

\section{Acknowledgments}
The work at Stanford and SLAC was supported by the U.S. Department of Energy (DOE), Office of Science, Basic Energy Sciences, Materials Sciences and Engineering Division, under Contract No. DE-AC02-76SF00515. ChemMatCARS Sector 15 is principally supported by the Divisions of Chemistry (CHE) and Materials Research (DMR), National Science Foundation, under grant number NSF/CHE-1346572.  Use of the PILATUS3 X CdTe 1M detector is supported by the National Science Foundation under the grant number NSF/DMR-1531283. Use of the Advanced Photon Source at Argonne National Laboratory was supported by the U. S. Department of Energy, Office of Science, Office of Basic Energy Sciences, under Contract No. DE-AC02-06CH11357. R.S. was supported by the Department of Defense (DoD) through the National Defense Science \& Engineering Graduate Fellowship (NDSEG) Program as well as an NSF Graduate Research Fellowship (DGE-1656518). We thank S. Lapidus for assistance at 11-BM, Y.-S. Chen and S.Y. Wang for assistance at 15-ID, and H.I. Karunadasa for generous access to equipment. Part of this work was performed at the Stanford Nano Shared Facilities (SNSF), supported by the NSF under award ECCS-1542152.

\bibliography{mendeley}

\begin{thebibliography}{42}
\expandafter\ifx\csname natexlab\endcsname\relax\def\natexlab#1{#1}\fi
\providecommand{\url}[1]{\texttt{#1}}
\providecommand{\href}[2]{#2}
\providecommand{\path}[1]{#1}
\providecommand{\DOIprefix}{doi:}
\providecommand{\ArXivprefix}{arXiv:}
\providecommand{\URLprefix}{URL: }
\providecommand{\Pubmedprefix}{pmid:}
\providecommand{\doi}[1]{\href{http://dx.doi.org/#1}{\path{#1}}}
\providecommand{\Pubmed}[1]{\href{pmid:#1}{\path{#1}}}
\providecommand{\bibinfo}[2]{#2}
\ifx\xfnm\relax \def\xfnm[#1]{\unskip,\space#1}\fi
\bibitem[{Balents(2010)}]{Balents2010}
\bibinfo{author}{L.~Balents}, \bibinfo{journal}{Nature} \bibinfo{volume}{464}
  (\bibinfo{year}{2010}) \bibinfo{pages}{199--208}.
  \DOIprefix\doi{10.1038/nature08917}.
\bibitem[{Norman(2016)}]{Norman2016}
\bibinfo{author}{M.~R. Norman}, \bibinfo{journal}{Rev. Mod. Phys.}
  \bibinfo{volume}{88} (\bibinfo{year}{2016}) \bibinfo{pages}{041002}.
  \DOIprefix\doi{10.1103/RevModPhys.88.041002}.
\bibitem[{Anderson(1973)}]{Anderson1973}
\bibinfo{author}{P.~W. Anderson}, \bibinfo{journal}{Mater. Res. Bull.}
  \bibinfo{volume}{8} (\bibinfo{year}{1973}) \bibinfo{pages}{153--160}.
  \DOIprefix\doi{10.1016/0025-5408(73)90167-0}.
\bibitem[{Ioffe et~al.(2002)Ioffe, Feigel'man, Ioselevich, Ivanov, Troyer, and
  Blatter}]{Ioffe2002}
\bibinfo{author}{L.~B. Ioffe}, \bibinfo{author}{M.~V. Feigel'man},
  \bibinfo{author}{A.~Ioselevich}, \bibinfo{author}{D.~Ivanov},
  \bibinfo{author}{M.~Troyer}, \bibinfo{author}{G.~Blatter},
  \bibinfo{journal}{Nature} \bibinfo{volume}{415} (\bibinfo{year}{2002})
  \bibinfo{pages}{503--506}. \DOIprefix\doi{10.1038/415503a}.
\bibitem[{Anderson(1987)}]{Anderson1987}
\bibinfo{author}{P.~W. Anderson}, \bibinfo{journal}{Science}
  \bibinfo{volume}{235} (\bibinfo{year}{1987}) \bibinfo{pages}{1196--1198}.
  \DOIprefix\doi{10.1126/science.235.4793.1196}.
\bibitem[{Savary and Balents(2017)}]{Savary2017}
\bibinfo{author}{L.~Savary}, \bibinfo{author}{L.~Balents},
  \bibinfo{journal}{Rep. Prog. Phys.} \bibinfo{volume}{80}
  (\bibinfo{year}{2017}) \bibinfo{pages}{016502}.
  \DOIprefix\doi{10.1088/0034-4885/80/1/016502}.
\bibitem[{Braithwaite et~al.(2004)Braithwaite, Mereiter, Paar, and
  Clark}]{Braithwaite2004}
\bibinfo{author}{R.~S.~W. Braithwaite}, \bibinfo{author}{K.~Mereiter},
  \bibinfo{author}{W.~H. Paar}, \bibinfo{author}{A.~M. Clark},
  \bibinfo{journal}{Mineral. Mag.} \bibinfo{volume}{68} (\bibinfo{year}{2004})
  \bibinfo{pages}{527--539}. \DOIprefix\doi{10.1180/0026461046830204}.
\bibitem[{Shores et~al.(2005)Shores, Nytko, Bartlett, and Nocera}]{Shores2005}
\bibinfo{author}{M.~P. Shores}, \bibinfo{author}{E.~A. Nytko},
  \bibinfo{author}{B.~M. Bartlett}, \bibinfo{author}{D.~G. Nocera},
  \bibinfo{journal}{J. Am. Chem. Soc.} \bibinfo{volume}{127}
  (\bibinfo{year}{2005}) \bibinfo{pages}{13462--13463}.
  \DOIprefix\doi{10.1021/ja053891p}.
\bibitem[{Han et~al.(2012)Han, Helton, Chu, Nocera, Rodriguez-Rivera, Broholm,
  and Lee}]{Han2012}
\bibinfo{author}{T.-H. Han}, \bibinfo{author}{J.~S. Helton},
  \bibinfo{author}{S.~Chu}, \bibinfo{author}{D.~G. Nocera},
  \bibinfo{author}{J.~A. Rodriguez-Rivera}, \bibinfo{author}{C.~Broholm},
  \bibinfo{author}{Y.~S. Lee}, \bibinfo{journal}{Nature} \bibinfo{volume}{492}
  (\bibinfo{year}{2012}) \bibinfo{pages}{406--10}.
  \DOIprefix\doi{10.1038/nature11659}.
\bibitem[{Fu et~al.(2015)Fu, Imai, Han, and Lee}]{Fu2015}
\bibinfo{author}{M.~Fu}, \bibinfo{author}{T.~Imai}, \bibinfo{author}{T.-H.
  Han}, \bibinfo{author}{Y.~S. Lee}, \bibinfo{journal}{Science}
  \bibinfo{volume}{350} (\bibinfo{year}{2015}) \bibinfo{pages}{655--658}.
  \DOIprefix\doi{10.1126/science.aab2120}.
\bibitem[{{de Vries} et~al.(2012){de Vries}, Wulferding, Lemmens, Lord,
  Harrison, Bonville, Bert, and Mendels}]{DeVries2012}
\bibinfo{author}{M.~A. {de Vries}}, \bibinfo{author}{D.~Wulferding},
  \bibinfo{author}{P.~Lemmens}, \bibinfo{author}{J.~S. Lord},
  \bibinfo{author}{A.~Harrison}, \bibinfo{author}{P.~Bonville},
  \bibinfo{author}{F.~Bert}, \bibinfo{author}{P.~Mendels},
  \bibinfo{journal}{Phys. Rev. B} \bibinfo{volume}{85} (\bibinfo{year}{2012})
  \bibinfo{pages}{014422}. \DOIprefix\doi{10.1103/PhysRevB.85.014422}.
\bibitem[{Han et~al.(2016)Han, Norman, Wen, Rodriguez-Rivera, Helton, Broholm,
  and Lee}]{Han2016b}
\bibinfo{author}{T.-H. Han}, \bibinfo{author}{M.~R. Norman},
  \bibinfo{author}{J.-J. Wen}, \bibinfo{author}{J.~A. Rodriguez-Rivera},
  \bibinfo{author}{J.~S. Helton}, \bibinfo{author}{C.~Broholm},
  \bibinfo{author}{Y.~S. Lee}, \bibinfo{journal}{Phys. Rev. B}
  \bibinfo{volume}{94} (\bibinfo{year}{2016}) \bibinfo{pages}{060409(R)}.
  \DOIprefix\doi{10.1103/PhysRevB.94.060409}.
\bibitem[{Elliott et~al.(2014)Elliott, Cooper, and Pring}]{Elliott2014}
\bibinfo{author}{P.~Elliott}, \bibinfo{author}{M.~A. Cooper},
  \bibinfo{author}{A.~Pring}, \bibinfo{journal}{Mineral. Mag.}
  \bibinfo{volume}{78} (\bibinfo{year}{2014}) \bibinfo{pages}{1755--1762}.
  \DOIprefix\doi{10.1180/minmag.2014.078.7.17}.
\bibitem[{Han et~al.(2014)Han, Singleton, and Schlueter}]{Han2014}
\bibinfo{author}{T.-H. Han}, \bibinfo{author}{J.~Singleton},
  \bibinfo{author}{J.~A. Schlueter}, \bibinfo{journal}{Phys. Rev. Lett.}
  \bibinfo{volume}{113} (\bibinfo{year}{2014}) \bibinfo{pages}{227203}.
  \DOIprefix\doi{10.1103/PhysRevLett.113.227203}.
\bibitem[{Jeschke et~al.(2015)Jeschke, Salvat-Pujol, Gati, Hoang, Wolf, Lang,
  Schlueter, and Valent\'i}]{Jeschke2015}
\bibinfo{author}{H.~O. Jeschke}, \bibinfo{author}{F.~Salvat-Pujol},
  \bibinfo{author}{E.~Gati}, \bibinfo{author}{N.~H. Hoang},
  \bibinfo{author}{B.~Wolf}, \bibinfo{author}{M.~Lang}, \bibinfo{author}{J.~A.
  Schlueter}, \bibinfo{author}{R.~Valent\'i}, \bibinfo{journal}{Phys. Rev. B}
  \bibinfo{volume}{92} (\bibinfo{year}{2015}) \bibinfo{pages}{094417}.
  \DOIprefix\doi{10.1103/PhysRevB.92.094417}.
\bibitem[{Guterding et~al.(2016)Guterding, Valent\'i, and
  Jeschke}]{Guterding2016a}
\bibinfo{author}{D.~Guterding}, \bibinfo{author}{R.~Valent\'i},
  \bibinfo{author}{H.~O. Jeschke}, \bibinfo{journal}{Phys. Rev. B}
  \bibinfo{volume}{94} (\bibinfo{year}{2016}) \bibinfo{pages}{125136}.
  \DOIprefix\doi{10.1103/PhysRevB.94.125136}.
\bibitem[{Liu et~al.(2015)Liu, Zou, Mei, and Liu}]{Liu2015a}
\bibinfo{author}{Z.~Liu}, \bibinfo{author}{X.~Zou}, \bibinfo{author}{J.-W.
  Mei}, \bibinfo{author}{F.~Liu}, \bibinfo{journal}{Phys. Rev. B}
  \bibinfo{volume}{92} (\bibinfo{year}{2015}) \bibinfo{pages}{220102(R)}.
  \DOIprefix\doi{10.1103/PhysRevB.92.220102}.
\bibitem[{Han et~al.(2016)Han, Isaacs, Schlueter, and Singleton}]{Han2016}
\bibinfo{author}{T.-H. Han}, \bibinfo{author}{E.~D. Isaacs},
  \bibinfo{author}{J.~A. Schlueter}, \bibinfo{author}{J.~Singleton},
  \bibinfo{journal}{Phys. Rev. B} \bibinfo{volume}{93} (\bibinfo{year}{2016})
  \bibinfo{pages}{214416}. \DOIprefix\doi{10.1103/PhysRevB.93.214416}.
\bibitem[{Pasco et~al.(2018)Pasco, Trump, Tran, Kelly, Hoffmann, Heinmaa,
  Stern, and McQueen}]{Pasco2018}
\bibinfo{author}{C.~M. Pasco}, \bibinfo{author}{B.~A. Trump},
  \bibinfo{author}{T.~T. Tran}, \bibinfo{author}{Z.~A. Kelly},
  \bibinfo{author}{C.~Hoffmann}, \bibinfo{author}{I.~Heinmaa},
  \bibinfo{author}{R.~Stern}, \bibinfo{author}{T.~M. McQueen},
  \bibinfo{journal}{Phys. Rev. Mat.} \bibinfo{volume}{2} (\bibinfo{year}{2018})
  \bibinfo{pages}{044406}. \URLprefix
  \url{https://doi.org/10.1103/PhysRevMaterials.2.044406}.
\bibitem[{{Bruker AXS Software Inc.}(2016)}]{BrukerAXSSoftwareInc2016}
\bibinfo{author}{{Bruker AXS Software Inc.}}, \bibinfo{howpublished}{{Madison,
  Wisconsin}}, \bibinfo{year}{2016}.
\bibitem[{Sheldrick(2015)}]{sheldrick2015}
\bibinfo{author}{G.~M. Sheldrick}, \bibinfo{journal}{Acta Crystallogr., Sect.
  C} \bibinfo{volume}{71} (\bibinfo{year}{2015}) \bibinfo{pages}{3--8}.
  \DOIprefix\doi{10.1107/S2053229614024218}.
\bibitem[{Dolomanov et~al.(2009)Dolomanov, Bourhis, Gildea, Howard, and
  Puschmann}]{Dolomanov2009}
\bibinfo{author}{O.~V. Dolomanov}, \bibinfo{author}{L.~J. Bourhis},
  \bibinfo{author}{R.~J. Gildea}, \bibinfo{author}{J.~A.~K. Howard},
  \bibinfo{author}{H.~Puschmann}, \bibinfo{journal}{J. Appl. Crystallogr.}
  \bibinfo{volume}{42} (\bibinfo{year}{2009}) \bibinfo{pages}{339--341}.
  \DOIprefix\doi{10.1107/S0021889808042726}.
\bibitem[{Toby and {Von Dreele}(2013)}]{toby2013}
\bibinfo{author}{B.~H. Toby}, \bibinfo{author}{R.~B. {Von Dreele}},
  \bibinfo{journal}{J. Appl. Crystallogr.} \bibinfo{volume}{46}
  (\bibinfo{year}{2013}) \bibinfo{pages}{544--549}.
  \DOIprefix\doi{10.1107/S0021889813003531}.
\bibitem[{Appelman(1969)}]{Appelman1969}
\bibinfo{author}{E.~H. Appelman}, \bibinfo{journal}{Inorg. Chem.}
  \bibinfo{volume}{8} (\bibinfo{year}{1969}) \bibinfo{pages}{223--227}.
  \DOIprefix\doi{10.1021/ic50072a008}.
\bibitem[{Feng et~al.(2017{\natexlab{a}})Feng, Wei, Liu, Yan, Wang, Luo,
  Senyshyn, dela Cruz, Yi, Mei, Meng, Shi, and Li}]{Feng2017a}
\bibinfo{author}{Z.~Feng}, \bibinfo{author}{Y.~Wei}, \bibinfo{author}{R.~Liu},
  \bibinfo{author}{D.~Yan}, \bibinfo{author}{Y.-C. Wang},
  \bibinfo{author}{J.~Luo}, \bibinfo{author}{A.~Senyshyn},
  \bibinfo{author}{C.~dela Cruz}, \bibinfo{author}{W.~Yi},
  \bibinfo{author}{J.-W. Mei}, \bibinfo{author}{Z.~Y. Meng},
  \bibinfo{author}{Y.~Shi}, \bibinfo{author}{S.~Li},
  \bibinfo{journal}{arXiv:1712.06732}  (\bibinfo{year}{2017}{\natexlab{a}}).
\bibitem[{Feng et~al.(2017{\natexlab{b}})Feng, Li, Meng, Yi, Wei, Zhang, Wang,
  Jiang, Liu, Li, Liu, Luo, Li, Zheng, Meng, Mei, and Shi}]{Feng2017}
\bibinfo{author}{Z.~Feng}, \bibinfo{author}{Z.~Li}, \bibinfo{author}{X.~Meng},
  \bibinfo{author}{W.~Yi}, \bibinfo{author}{Y.~Wei},
  \bibinfo{author}{J.~Zhang}, \bibinfo{author}{Y.-C. Wang},
  \bibinfo{author}{W.~Jiang}, \bibinfo{author}{Z.~Liu},
  \bibinfo{author}{S.~Li}, \bibinfo{author}{F.~Liu}, \bibinfo{author}{J.~Luo},
  \bibinfo{author}{S.~Li}, \bibinfo{author}{G.-q. Zheng},
  \bibinfo{author}{Z.~Y. Meng}, \bibinfo{author}{J.-W. Mei},
  \bibinfo{author}{Y.~Shi}, \bibinfo{journal}{Chin. Phys. Lett.}
  \bibinfo{volume}{34} (\bibinfo{year}{2017}{\natexlab{b}})
  \bibinfo{pages}{077502}. \DOIprefix\doi{10.1088/0256-307X/34/7/077502}.
\bibitem[{Chu et~al.(2011)Chu, M{\"{u}}ller, Nocera, and Lee}]{Chu2011}
\bibinfo{author}{S.~Chu}, \bibinfo{author}{P.~M{\"{u}}ller},
  \bibinfo{author}{D.~G. Nocera}, \bibinfo{author}{Y.~S. Lee},
  \bibinfo{journal}{Appl. Phys. Lett.} \bibinfo{volume}{98}
  (\bibinfo{year}{2011}) \bibinfo{pages}{092508}.
  \DOIprefix\doi{10.1063/1.3562010}.
\bibitem[{Han et~al.(2011)Han, Helton, Chu, Prodi, Singh, Mazzoli,
  M{\"{u}}ller, Nocera, and Lee}]{Han2011}
\bibinfo{author}{T.~H. Han}, \bibinfo{author}{J.~S. Helton},
  \bibinfo{author}{S.~Chu}, \bibinfo{author}{A.~Prodi}, \bibinfo{author}{D.~K.
  Singh}, \bibinfo{author}{C.~Mazzoli}, \bibinfo{author}{P.~M{\"{u}}ller},
  \bibinfo{author}{D.~G. Nocera}, \bibinfo{author}{Y.~S. Lee},
  \bibinfo{journal}{Phys. Rev. B} \bibinfo{volume}{83} (\bibinfo{year}{2011})
  \bibinfo{pages}{100402(R)}. \DOIprefix\doi{10.1103/PhysRevB.83.100402}.
\bibitem[{sup(????)}]{supp}
???? \bibinfo{note}{See Supplementary Material at XX for Rietveld refinements,
  structure solution and refinement information, selected bond lengths, IR
  spectra, TGA data and analysis, and Crystallographic Information Files
  (CIFs).}
\bibitem[{Momma and Izumi(2011)}]{Momma2011}
\bibinfo{author}{K.~Momma}, \bibinfo{author}{F.~Izumi}, \bibinfo{journal}{J.
  Appl. Crystallogr.} \bibinfo{volume}{44} (\bibinfo{year}{2011})
  \bibinfo{pages}{1272--1276}. \DOIprefix\doi{10.1107/S0021889811038970}.
\bibitem[{Burns et~al.(1995)Burns, Cooper, and Hawthorne}]{Burns1995a}
\bibinfo{author}{P.~C. Burns}, \bibinfo{author}{M.~A. Cooper},
  \bibinfo{author}{F.~C. Hawthorne}, \bibinfo{journal}{Can. Mineral.}
  \bibinfo{volume}{33} (\bibinfo{year}{1995}) \bibinfo{pages}{633--639}.
\bibitem[{Fanfani et~al.(1973)Fanfani, Nunzi, Zanazzi, and
  Zanzari}]{Fanfani1973}
\bibinfo{author}{L.~Fanfani}, \bibinfo{author}{A.~Nunzi},
  \bibinfo{author}{P.~F. Zanazzi}, \bibinfo{author}{A.~R. Zanzari},
  \bibinfo{journal}{Mineral. Mag.} \bibinfo{volume}{39} (\bibinfo{year}{1973})
  \bibinfo{pages}{264--270}. \DOIprefix\doi{10.1180/minmag.1973.039.303.02}.
\bibitem[{McLean and Anthony(1972)}]{Mclean1972}
\bibinfo{author}{W.~J. McLean}, \bibinfo{author}{J.~W. Anthony},
  \bibinfo{journal}{Am. Mineral.} \bibinfo{volume}{57} (\bibinfo{year}{1972})
  \bibinfo{pages}{426--438}.
\bibitem[{Freedman et~al.(2010)Freedman, Han, Prodi, M{\"{u}}ller, Huang, Chen,
  Webb, Lee, McQueen, and Nocera}]{Freedman2010}
\bibinfo{author}{D.~E. Freedman}, \bibinfo{author}{T.~H. Han},
  \bibinfo{author}{A.~Prodi}, \bibinfo{author}{P.~M{\"{u}}ller},
  \bibinfo{author}{Q.-Z. Huang}, \bibinfo{author}{Y.-S. Chen},
  \bibinfo{author}{S.~M. Webb}, \bibinfo{author}{Y.~S. Lee},
  \bibinfo{author}{T.~M. McQueen}, \bibinfo{author}{D.~G. Nocera},
  \bibinfo{journal}{J. Am. Chem. Soc.} \bibinfo{volume}{132}
  (\bibinfo{year}{2010}) \bibinfo{pages}{16185--16190}.
  \DOIprefix\doi{10.1021/ja1070398}.
\bibitem[{Burns and Hawthorne(1996)}]{Burns1996}
\bibinfo{author}{P.~C. Burns}, \bibinfo{author}{F.~C. Hawthorne},
  \bibinfo{journal}{Can. Mineral.} \bibinfo{volume}{34} (\bibinfo{year}{1996})
  \bibinfo{pages}{1089--1105}.
\bibitem[{Schuiskii and Zorina(2013)}]{Schuiskii2013}
\bibinfo{author}{A.~V. Schuiskii}, \bibinfo{author}{M.~L. Zorina},
  \bibinfo{journal}{J. Appl. Spectrosc.} \bibinfo{volume}{80}
  (\bibinfo{year}{2013}) \bibinfo{pages}{576--580}.
  \DOIprefix\doi{10.1007/s10812-013-9808-2}.
\bibitem[{Sithole et~al.(2012)Sithole, Ngom, Khamlich, Manikanadan, Manyala,
  Saboungi, Knoessen, Nemutudi, and Maaza}]{Sithole2012}
\bibinfo{author}{J.~Sithole}, \bibinfo{author}{B.~D. Ngom},
  \bibinfo{author}{S.~Khamlich}, \bibinfo{author}{E.~Manikanadan},
  \bibinfo{author}{N.~Manyala}, \bibinfo{author}{M.~L. Saboungi},
  \bibinfo{author}{D.~Knoessen}, \bibinfo{author}{R.~Nemutudi},
  \bibinfo{author}{M.~Maaza}, \bibinfo{journal}{Appl. Surf. Sci.}
  \bibinfo{volume}{258} (\bibinfo{year}{2012}) \bibinfo{pages}{7839--7843}.
  \DOIprefix\doi{10.1016/j.apsusc.2012.04.073}.
\bibitem[{Domenicali(1950)}]{Domenicali1950}
\bibinfo{author}{C.~A. Domenicali}, \bibinfo{journal}{Phys. Rev.}
  \bibinfo{volume}{78} (\bibinfo{year}{1950}) \bibinfo{pages}{458--467}.
  \DOIprefix\doi{10.1103/PhysRev.78.458}.
\bibitem[{Ramirez(1994)}]{Ramirez1994}
\bibinfo{author}{A.~P. Ramirez}, \bibinfo{journal}{Annu. Rev. Mater. Sci.}
  \bibinfo{volume}{24} (\bibinfo{year}{1994}) \bibinfo{pages}{453--480}.
  \DOIprefix\doi{10.1146/annurev.ms.24.080194.002321}.
\bibitem[{Schock(1985)}]{Schock1985}
\bibinfo{editor}{R.~N. Schock} (Ed.), \bibinfo{title}{{Point Defects in
  Minerals}}, \bibinfo{publisher}{American Geophysical Union},
  \bibinfo{year}{1985}. \DOIprefix\doi{10.1029/GM031}.
\bibitem[{Hobbs(1984)}]{Hobbs1984}
\bibinfo{author}{B.~E. Hobbs}, \bibinfo{journal}{J. Geophys. Res.}
  \bibinfo{volume}{89} (\bibinfo{year}{1984}) \bibinfo{pages}{4026--2038}.
  \DOIprefix\doi{10.1029/JB089iB06p04026}.
\bibitem[{Burns and Hawthorne(1995)}]{Burns1995}
\bibinfo{author}{P.~C. Burns}, \bibinfo{author}{F.~C. Hawthorne},
  \bibinfo{journal}{Can. Mineral.} \bibinfo{volume}{33} (\bibinfo{year}{1995})
  \bibinfo{pages}{889--905}.

\end{thebibliography}

\end{document}